\documentclass[12pt]{article}%
\pdfoutput=1
\usepackage[nosort]{cite}
\usepackage{graphicx}
\usepackage{multicol}
\usepackage{amsfonts}
\usepackage{amsthm}
\usepackage{amssymb}
\usepackage{amsmath}
\usepackage{singleauthor}
\usepackage{afterpage}
\usepackage{setspace}
\usepackage{verbatim}
\usepackage{slashed}
\usepackage{color}
\usepackage{longtable}
\usepackage{float}
\usepackage{subcaption}
\usepackage{epsfig}
\usepackage{bm}
\usepackage{enumerate}
\usepackage{epstopdf}
\usepackage[enableskew, vcentermath]{youngtab}
\usepackage{adjustbox}
\newsavebox{\mysavebox}

\usepackage[margin=1in]{geometry}
\usepackage{titletoc}%
\usepackage{hyperref}

\newtheoremstyle{named}{}{}{\itshape}{}{\bfseries}{.}{.5em}{#3}
\theoremstyle{named}
\newtheorem*{namedconjecture}{Conjecture}

\newcommand{\ba}{\begin{eqnarray}}
\newcommand{\ea}{\end{eqnarray}}

\newcommand{\cF}{\mathcal{F}}

\newcommand{\cI}{\mathcal{I}}

\newcommand{\cM}{\mathcal M}

\newcommand{\Mpl}{M_{\rm Pl}}
\newcommand{\Mpld}{M_{\rm Pl;d}}
\newcommand{\Mten}{M_{\rm Pl;10}}
\newcommand{\Mnine}{M_{\rm Pl;9}}
\newcommand{\Mfive}{M_{\rm Pl;5}}

\newcommand{\mKK}{m_{\textrm{KK}}}
\newcommand{\mwind}{m_{\textrm{wind}}}
\newcommand{\mDwind}{\tilde{m}_{\textrm{wind}}}
\newcommand{\Mstring}{M_{\textrm{string}}}

\def\be{\begin{equation}}
\def\ee{\end{equation}}

\makeatletter \@addtoreset{equation}{section} \makeatother

\begin{document}

\date{\today}

\title{Revisiting the Refined Distance Conjecture}

\institution{BERKELEY}{Physics Department, University of California, Berkeley CA 94720 USA}

\authors{Tom Rudelius\worksat{\BERKELEY}\footnote{e-mail: {\tt rudelius@berkeley.edu}}}

\abstract{
The Distance Conjecture of Ooguri and Vafa holds that any infinite-distance limit in the moduli space of a quantum gravity theory must be accompanied by a tower of exponentially light particles, which places tight constraints on the low-energy effective field theories in these limits. One attempt to extend these constraints to the interior of moduli space is the refined Distance Conjecture, which holds that the towers of light particles predicted by the Distance Conjecture must appear any time a modulus makes a super-Planckian excursion in moduli space. In this note, however, we point out that a tower which satisfies the Distance Conjecture in an infinite-distance limit of moduli space may be parametrically heavier than the Planck scale for an arbitrarily long geodesic distance. This means that the refined Distance Conjecture, in its most naive form, does not place meaningful constraints on low-energy effective field theory. This motivates alternative refinements of the Distance Conjecture, which place an absolute upper bound on the tower mass scale in the interior of moduli space. We explore two possibilities, providing evidence for them and briefly discussing their implications.
}

\maketitle


\enlargethispage{\baselineskip}

\setcounter{tocdepth}{2}

\section{Introduction}\label{sec:INTRO}

It is well known that when it comes to conjectures about quantum gravity, rigor is inversely proportional to relevance. Statements that are supported by compelling evidence--such as the absence of global symmetries and the mild form of the Weak Gravity Conjecture \cite{ArkaniHamed:2006dz}--tend to place very weak constraints on testable low-energy physics. Meanwhile, conjectures that propose tight phenomenological constraints tend to be much more speculative.
 
One of the more rigorously established features of quantum gravity is the presence of exponentially light towers of particles in infinite-distance limits of moduli space. This is encoded in the Distance Conjecture of Ooguri and Vafa \cite{Ooguri:2006in}:
\vspace{.2cm}
           \begin{namedconjecture}[The Distance Conjecture]
Let $\cM$ be the scalar field moduli space of a quantum gravity theory in $d \geq 4$ dimensions, parametrized by vacuum expectation values of massless scalar fields. Compared to the theory at some point $p_0 \in \mathcal{M}$, the theory at a point $p \in \mathcal{M}$ has an infinite tower of particles, each with mass scaling as
\be
m(p) \sim m(p_0) \exp( -\lambda ||p - p_0|| )\,,
\label{DCdef}
\ee 
where $||p - p_0||$ is the geodesic distance in $\mathcal{M}$ between $p$ and $p_0$, and $\lambda$ is some order-one number in Planck units $(8 \pi G = \kappa_d^2 = \Mpld^{2-d} = 1)$.
            \end{namedconjecture}
    \vspace{.1cm}
\noindent

As noted, the Distance Conjecture is well-supported by examples in string theory \cite{Baume:2016psm, Klaewer:2016kiy, Blumenhagen:2018nts,Corvilain:2018lgw, Joshi:2019nzi, Erkinger:2019umg, Gendler:2020dfp} and conformal field theory \cite{Perlmutter:2020buo, Baume:2020dqd}. However, the price one pays for this relative rigor is a dearth of experimental consequences. Likely the most promising route for connecting the Distance Conjecture to observable physics is provided by large-field inflation, which involves a scalar field that rolls a super-Planckian distance in scalar field space. This smells like the sort of thing that the Distance Conjecture might be able to constrain, but there are several issues that must be addressed before the Distance Conjecture can give meaningful constraints on inflation.

One of these issues\footnote{For a discussion of other issues regarding applications of the Distance Conjecture to inflation, see Section \ref{CONC}.} is the fact that the Distance Conjecture is primarily concerned with infinite-distance limits of moduli space. Inflation does not require an infinite field traversal, so it is not clear that constraints on the infinite-distance limits of field space translate into meaningful constraints on inflation. Relatedly, many models of inflation involve periodic scalar fields (i.e., axions), which have compact support and thus are not immediately subject to constraints on infinite-distance limits in moduli space. 

One popular proposal for handling this issue goes by the name of the refined Distance Conjecture \cite{Baume:2016psm, Klaewer:2016kiy}, which holds that the exponential scaling behavior of the towers predicted by the Distance Conjecture is supposed to set in within an order-one distance in field space. This could ostensibly limit the length of an inflationary field traversal that can occur before the light towers come into play.

However, this refinement of the conjecture suffers an important drawback: even if the mass $m(p)$ of a tower of particles at a point $p$ is exponentially suppressed relative to the mass $m(p_0)$ at a point $p_0$ in accordance with \eqref{DCdef}, there is no guarantee that the mass $m(p)$ will be suppressed relative to the Planck scale. In other words, the mass scales $m(p_0)$ and $m(p)$ might \emph{both} be parametrically above the Planck scale, in which case the conjecture places no interesting constraints on low-energy physics in this part of moduli space.

This issue is not merely a pathological possibility, but rather it arises even in supersymmetric string theory in ten dimensions. In any such theory, there is a tower of string oscillator modes beginning at the string scale,
\begin{equation}
M_s \sim \Mten \exp( - \frac{\phi}{\sqrt{8}} ) \,,
\end{equation}
where $\phi \equiv - \log (g_s)/\sqrt{2}$ is the canonically normalized dilaton. Setting $\phi(p) = 0$, taking $\phi(p_0) \ll 0$, and holding any other moduli fixed, we find that the (refined) Distance Conjecture is indeed satisfied by the string oscillator modes, but these modes are super-Planckian everywhere along the geodesic between $p_0$ and $p$.



It is worth noting, however, that this example features not only a tower of string oscillator modes, which become light in the limit $\phi \rightarrow \infty$, but also a tower of modes that become light in the $\phi \rightarrow -\infty$ limit, the nature of which depends on the string theory in question. It is this latter tower, rather than the heavier tower of Kaluza-Klein modes, which plays a role in the low-energy effective field theories along the geodesic between $p_0$ and $p$.

In light of this example, it is tempting to define an alternative, bidirectional refinement of the Distance Conjecture, which demands that \eqref{DCdef} should be satisfied with $m(p), m(p_0) < \Mpld$ but permits a switch of $p$ and $p_0$, reversing the orientation of the geodesic. However, as we illustrate with an example in the following section, this naive refinement also runs into trouble: there exist geodesics $\gamma(s)$ in moduli spaces of quantum gravity theories for which one tower of particles becomes light as $s \rightarrow \infty$, another becomes light as $s \rightarrow -\infty$, and yet both towers are parametrically heavier than the Planck scale in an arbitrarily large intermediate regime of finite $s$.


However, we will see that these examples feature additional towers, so that at any point in moduli space at least one of the towers will be at or below the Planck scale.
This motivates a pair of closely related conjectures, which may be viewed as alternative refinements of the Distance Conjecture:

\vspace{.2cm}
           \begin{namedconjecture}[Conjecture 1]
Consider the collection of towers $\cI$ which satisfy the Distance Conjecture in the various infinite-distance limits of moduli space $\cM$, and let $m_i(p)$ be the mass scale of the $i^\text{th}$ such tower at the point $p$. Then, for all $p \in \cM$, there exists $i \in \cI$ such that 
\begin{equation}
m_i(p) \leq c_1 \Mpld\,,
\label{ourbound}
\end{equation}
where $c_1$ is an order-one constant that remains to be determined.
            \end{namedconjecture}
\noindent

\vspace{.2cm}
           \begin{namedconjecture}[Conjecture 2]
Let $\cM$ be the scalar field moduli space of a quantum gravity theory and let $\gamma(p_0, p)$ be a path with endpoints $p,p_0 \in \cM$. Then, there exists a point $p_1$ on the path $\gamma(p_0, p)$ such that the theory at $p_1$ has an infinite tower of particles beginning at the mass scale
\begin{equation}
m(p_1) \leq c_2 \Mpld \exp (  - \lambda || p - p_0 ||/2 )\,,
\label{ourbound2}
\end{equation}
where $|| p - p_0||$ is the geodesic distance between $p$ and $p_0$,\footnote{Note that $||p-p_0||$ is the length of the \emph{shortest} geodesic between $p_0$ and $p$, not necessarily the length of the path $\gamma(p_0,p)$. This is especially important in cases where the moduli space is given by a discrete quotient of some larger covering space, as it allows us to restrict our choice of representatives of $p$ and $p_0$ to lie within a single fundamental domain. See \S\ref{ssec:10d} for an example of this. \label{footdef}} and $c_2$ and $\lambda$ are order-one constants that remain to be determined.
            \end{namedconjecture}
\noindent

Note that conjecture 2 is very similar to conjectures made previously in \cite{Hebecker:2017lxm}.\footnote{We thank Arthur Hebecker for bringing these conjectures to our attention after the first version of this paper was posted.}

In subsequent sections, we will provide evidence for these conjectures in simple examples.
These examples satisfy the bounds \eqref{ourbound}, \eqref{ourbound2} with $c_1, c_2 \lesssim  2 \pi$, though further work is needed to pin down universal upper bounds on these order-one coefficients. The recent work \cite{Etheredge:2022opl} argued that the coefficient $\lambda$ is bounded below as $\lambda \geq 1/\sqrt{d-2}$ in $d$ dimensions, and the examples we study below fall under the scope of that analysis.

These conjectures can be justified intuitively as follows: in quantum gravity, continuous parameters are controlled by vacuum expectation values (vevs) of scalar fields. Parametric growth of some mass scale $m_i \gg \Mpl$ therefore requires parametric growth of some scalar field, which is precisely when the Distance Conjecture demands the existence of some exponentially light tower. This suggests that there can never be one tower of particles that is parametrically heavier than the Planck scale unless another tower is parametrically lighter than the Planck scale.

This reasoning--and even the meaning of the two conjectures--is somewhat hazy for non-supersymmetric towers of states, as it may be difficult to define the mass scale of a tower which heavier than the Planck scale. Consequently, in this work we will focus primarily on towers of BPS particles, the mass scale of which remains well-defined even when it is super-Planckian. More generally, we might say that our conjectures simply demand the existence of a tower with a well-defined, sub-Planckian mass scale, so they cannot be satisfied by super-Planckian slop.

The two conjectures are closely related, but they are not quite equivalent to one another.\footnote{We thank Ben Heidenreich for his insights on these conjectures and the relationship between them.} By taking $||p -p_0|| \rightarrow 0$, conjecture 2 implies that at least one tower of particles cannot be parametrically heavier than the Planck scale at any point in moduli space. This tower will then satisfy conjecture 1 (with $c_1 = c_2$) provided the tower becomes exponentially light in some infinite-distance limit in moduli space. 

Conversely, if conjecture 1 is satisfied, then at the midpoint of the shortest geodesic between $p$ and $p_0$, at least one tower of particles will have a sub-Planckian mass scale. In many cases, this tower will decay exponentially along the geodesic in one direction or the other in accordance with the Distance Conjecture, and in turn it will satisfy conjecture 2 with $c_2 \approx c_1$. In the cases where this tower does not decay, we will argue that an additional, lighter tower is present, thereby ensuring that conjecture 2 is satisfied.

Conjecture 1 is somewhat easier to verify in practice, while conjecture 2 is more relevant for super-Planckian traversals of scalar fields, such as the ones that occur in models of large-field inflation. In particular, conjecture 2 places an upper bound on the mass scale of the lightest tower along the geodesic that decays exponentially with increasing geodesic distance. If this tower consists of either string oscillator modes or Kaluza-Klein modes in some duality frame (as suggested by the Emergent String Conjecture \cite{Lee:2019wij}), then this mass scale must lie below the Hubble scale in any model of inflation, $H < \mKK$, $\Mstring$. Conjecture 2 thus addresses one gap between the Distance Conjecture and inflation by extending the former to the interior of moduli space; additional gaps will be discussed in Section \ref{CONC}.


The remainder of the paper is structured as follows. In Section \ref{EX}, we show how our conjectures are satisfied in simple examples of string theory compactified on a torus, and we demonstate that other refinements of the Distance Conjecture are unable to constrain low-energy effective field theories. 
In Section \ref{5D}, we show how our conjectures are satisfied in two simple Calabi-Yau compactifications of M-theory to five dimensions.
In Section \ref{EMD}, we explain how conjecture 1 follows from the tower Weak Gravity Conjecture in a simple Einstein-Maxwell-Dilaton theory in four dimensions. In Section \ref{CONC}, we conclude this note with a brief discussion of implications and next steps.

\section{String Theory on $T^n$}\label{EX}

\subsection{String theory in ten dimensions}\label{ssec:10d}

String theory in ten dimensions has a tower of string oscillator modes beginning at the string scale,
\begin{equation}
M_s = \sqrt{2 \pi T} = \frac{1}{\ell_s}\,.
\end{equation}
The string scale is related to the Planck scale in ten dimensions as
\begin{equation}
\Mten = \left( \frac{4 \pi }{g_s^2}  \right)^{1/8} \frac{M_s}{2\pi}\,.
\end{equation}
The string oscillator modes become light (in Planck units) in the limit $g_s \equiv \exp(- \phi \sqrt{2} ) \rightarrow 0$, satisfying the Distance Conjecture with a coefficient $\lambda = 1/\sqrt{8}$.

Meanwhile, the strong coupling limit $g_s \rightarrow \infty$ also features a tower of light particles, the nature of which depends on the string theory in question. For Type IIA string theory or $E_8 \times E_8$ heterotic string theory, the strong coupling limit is a decompactification limit to M-theory, which features a tower of particles with 
\begin{equation}
\mKK = \frac{M_s}{g_s}\,.
\end{equation}
In the Type IIA case, these Kaluza-Klein modes are also known as D0-branes.

In the case of Type IIB string theory, Type I string theory, or $SO(32)$ heterotic string theory, on the other hand, the strong coupling limit represents the weak coupling limit of a dual string theory, which features a tower of string oscillator modes with 
\begin{equation}
\tilde M_s = \frac{M_s}{\sqrt{g_s}}\,.
\end{equation}

In any case, the energy scales of the towers for the weak/strong coupling limits coincide when $g_s = 1$, at which point
\begin{equation}
M_s = \tilde M_s = \mKK =  \frac{2 \pi \Mten}{(4 \pi)^{1/8}}\,.
\end{equation} 
At any other point in moduli space, at least one tower will be lighter than this value. This ensures that conjecture 1 is satisfied with $c_1 = 2 \pi  / (4 \pi)^{1/8} $.

For any value of the dilaton $\phi$, there is a tower of particles beginning at a scale $m \leq  2 \pi \Mten / (4 \pi)^{1/8}$. As the dilaton varies over a distance $\Delta \phi$, there will be at least one point $p_1$ with $|\phi(p_1)| \geq |\Delta \phi| /2$, and at this point one of the towers for the strong/weak coupling limits will have mass bounded above as
\begin{equation}
m(p_1) \leq  2 \pi \Mten / (4 \pi)^{1/8} \exp( - |\Delta \phi|/(4 \sqrt{2}))\,.
\end{equation}
Comparing with \eqref{ourbound2}, we see that conjecture 2 is satisfied along this geodesic with $ c_2 =  2 \pi  / (4 \pi)^{1/8} $, $\lambda  =1/\sqrt{d-2} = 1/\sqrt{8}$.

\begin{figure}
\begin{center}
\includegraphics[width = 90mm]{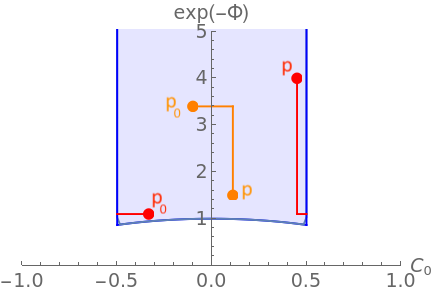} 
\end{center}
\caption{Paths in the fundamental domain of Type IIB moduli space. The fundamental domain of Type IIB moduli space is given by the (shaded blue) keyhole region $|\tau| \geq 1,$ $-1/2 \leq \Re(\tau) < 1/2$, with $\tau = C_0 + i \exp( -\Phi)$. The geodesic distance between any two points $p$, $p_0$ in the keyhole region is bounded above by the length of the path shown, one segment of which has fixed $\Phi$ and the other of which has fixed $C_0$.}
\label{keyhole}
\end{figure}

For Type IIB string theory, there is one remaining subtlety: we have not yet accounted for the axion field, $C_0$. To address this issue, we must use the $SL(2, \mathbb{Z})$ of Type IIB string theory, which implies that the fundamental domain of moduli space of Type IIB string theory is given by the famous keyhole of Figure \ref{keyhole}. 
Within the fundamental domain, the lightest tower of string modes is that of the fundamental string, with
\begin{equation}
M_s = \frac{2 \pi \Mten}{(4 \pi)^{1/8}} \exp(\Phi/4)\,,
\end{equation}
where $\Phi = -\sqrt{2} \phi = \log g_s$ is the conventionally normalized dilaton. Within the fundamental domain, we have the bound $\exp(\Phi) \leq 2/\sqrt{3}$, which implies that conjecture 1 is satisfied with $c_1 = 2 \pi  / (3 \pi)^{1/8}$. This is larger by a factor of $(4/3)^{1/8} \approx 1.04$ than the value of $c_1$ we found when we ignored the axion.

When it comes to conjecture 2, one could in principle choose the points $p$ and $p_0$ to lie in different fundamental domains of the $SL(2, \mathbb{Z})$ action, connected by the shortest geodesic between them within the larger covering space. This is gauge-equivalent to choosing both $p$ and $p_0$ to lie in the same fundamental domain but considering a non-minimal geodesic between them, which may traverse the width of the keyhole region many times before finally connecting the two points $p$ and $p_0$. However, as emphasized in footnote \ref{footdef} above, we have carefully defined Conjecture 2 so that the relevant geodesic distance $||p-p_0||$ is the length of the \emph{shortest} geodesic between $p$ and $p_0$. Thus, for placing an upper bound on the coefficient $c_2$ in conjecture 2, it suffices to choose $p$ and $p_0$ to each lie within the keyhole region and to bound the geodesic $||p - p_0||$ along the shortest geodesic between these points.

With this, the axion ultimately has only a small effect on the coefficient $c_2$ of conjecture 2, which is due to the fact that the points $p_0$ and $p$ may now differ not only by a change in the dilaton $\Phi$, but also by a change in the axion, $C_0$.
Within the fundamental domain, the geodesic distance between $p$ and $p_0$ is bounded above by
\begin{equation}
||p-p_0|| \leq \frac{1}{\sqrt{2}} |\Delta \Phi| + \frac{1}{\sqrt{2}} \exp(  \Phi ) |\Delta C_0| \leq \frac{1}{\sqrt{2}} (\log(2/\sqrt{3}) - \Phi)+ \frac{1}{\sqrt{6}}  \,,
\label{keyholebound}
\end{equation}
where in the first step we have used the path shown in Figure \ref{keyhole}, and in the last step we have used the fact that $\exp(\Phi) \leq 2/\sqrt{3}$ everywhere in the keyhole region, while $|\Delta C_0| \leq 1/2$.
Meanwhile, the mass scale associated with the lightest tower of string modes is given by 
\begin{equation}
m =   \frac{2 \pi \Mten }{ (4 \pi)^{1/8}} \exp(  \Phi / 4 ) \leq \frac{2 \pi \Mten }{ (3 \pi)^{1/8}}    \exp \left( \frac{1}{4\sqrt{3}} - \frac{ || p-p_0 ||}{ \sqrt{8}} \right)\,,
\end{equation}
where in the last step we have used \eqref{keyholebound}. We conclude that conjecture 2 is satisfied for Type IIB string theory in ten dimensions with $\lambda = 2 /\sqrt{8}$ and
\begin{equation}
c_2 = \frac{2 \pi \Mten }{ (3 \pi)^{1/8}}    \exp (1 / (4 \sqrt{3} ) )\,.
\end{equation}
This value of $c_2$ is larger by a factor of $(4/3)^{1/8} \exp(1/(4\sqrt{3})) \approx 1.2$ than the value of $c_2$ we obtained when we ignored the axions. The value of $\lambda$ we obtained here is twice that of the value we obtained above for Type IIA string theory, a result of the fact that the domain of $\phi$ is semi-infinite within the keyhole region.

More generally, the inclusion of the periodic directions of moduli space makes our conjectures a bit more difficult to verify explicitly, but we expect that the inclusion of these directions will introduce at most order-one corrections to the factors $c_1$, $c_2$.


\subsection{String theory on $S^1$}

Consider next the case of Type IIB string theory compactified on a circle of radius $R$. For simplicity, we set the axion to vanish, leaving a two-dimensional moduli space parametrized by the dilaton $\phi$ and the radion $\rho$.

The Distance Conjecture is satisfied in every direction in the dilaton-radion plane by one of three towers: Kaluza-Klein modes, winding modes of the fundamental string, and winding modes of the D-string. The mass scales associated with these three towers are given by:
\begin{equation}
\mKK = \frac{1}{R} \,,~~~~ \mwind = M_s^2 R \,,~~~~ \mDwind = \frac{M_s^2 R}{g_s}\,.
\end{equation}
Meanwhile, the Planck scale in nine dimensions is related to $M_s$, $R$, and $g_s$ as
\begin{equation}
\Mnine=  \left( \frac{8 \pi^2 R }{g_s^2}  \right)^{1/7} \left( \frac{M_s}{2\pi} \right)^{8/7}
\end{equation}

The minimum of $\mKK$, $\mwind$, and $\mDwind$ is maximized when the three energy scales are set equal to one another. This occurs when $R = M_s$, $g_s =1$, at which point
\begin{equation}
\mKK = \mwind = \mDwind = M_s =  \frac{2 \pi \Mnine}{(4 \pi)^{1/7}} \,.
\end{equation}
As we move away from this point, at least one of these three towers becomes lighter in Planck units. To see this, let us first define the $\zeta$-vector of a particle of mass $m$ at a given point in moduli space by
\begin{equation}
\zeta_i = - \frac{\partial}{\partial \phi^i} \log m\,.
\label{zeta}
\end{equation}
Then, we may write the Kaluza-Klein and winding scales in terms of the canonically normalized dilaton $\phi$ and radion $\rho$ in nine dimensions:
\begin{align}
\mKK = m_0 \exp(- \frac{\phi}{\sqrt{7}} - \rho)\,,~~~\mwind = m_0 \exp(- \frac{\phi}{\sqrt{7}} + \rho) \,, ~~~
\mDwind &=  m_0  \exp(\frac{5\phi}{2\sqrt{7}} + \frac{ \rho}{2})\,,
\label{scaling9d}
\end{align} 
where $m_0 = 2 \pi \Mnine / (4 \pi)^{1/7}$, $\phi = - \sqrt{7} \log(M_s/m_0)$, and $\rho = \log(R M_s)$.
With these conventions, $\rho \rightarrow \infty$ is the large volume limit, and $\phi \rightarrow \infty$ is the weak string coupling limit. This scaling behavior leads to $\zeta$-vectors
\begin{equation}
\vec{\zeta}^{\text{KK}}  = (\frac{1}{\sqrt{7}} , 1) \,,~~~\vec{\zeta}^{\text{wind}}  = (\frac{1}{\sqrt{7}} , -1)\,,~~~\vec{\zeta}^{\text{D-wind}}  = (-\frac{5}{2\sqrt{7}} , - \frac{1}{2})\,.
\end{equation}
It is straightforward to check that the convex hull of these $\zeta$-vectors contains the ball of radius $1/\sqrt{d-2}$, ensuring that the sharpened Distance Conjecture of \cite{Etheredge:2022opl} is satisfied in all infinite-distance limits of moduli space. This means that for any point $(\phi_0, \rho_0)$ in moduli space, at least one of the three $\zeta$-vectors satisfies $\zeta \cdot (\phi_0, \rho_0) \geq 1/\sqrt{d-2}$, which means that at least one of the three mass scales $\mKK$, $\mwind$, and $\mDwind$ will be suppressed by its exponent relative to $m_0$. This ensures that conjecture 1 is satisfied with $c_1 =  m_0 / \Mnine = 2 \pi / (4 \pi)^{1/7}$.

Any geodesic in the $\phi$-$\rho$ plane of length $L$ will then have at least one point $p_1$ that lies a distance of $L/2$ or greater from the origin. At such a point, the smallest of $\mKK$, $\mwind$, and $\mDwind$ will be bounded above as
\begin{equation}
m(p_1) \leq m_0 \exp( - L/(2 \sqrt{7}) ) 
\end{equation}
Thus, conjecture 2 is satisfied in nine dimensions, this time with coefficients $c_2 = 2 \pi / (4 \pi)^{1/7}$, $\lambda = 1/\sqrt{d-2} = 1/\sqrt{7}$.

What happens if we include the axions in our analysis? Giving a vev to an axion field does not affect the characteristic mass scales of the Kaluza-Klein tower or the tower of fundamental string winding modes, but it does have an order-one effect on the mass scale of D-string winding modes. Conjecture 1 will remain satisfied with an order-one shift of the constant $c_1$, as in the case of 10-dimensional Type IIB string theory discussed above. 

Similarly, the coefficient $c_2$ of conjecture 2 may be altered once we allow our geodesics to explore the periodic directions of moduli space. Using dualities, we may choose a fundamental domain of moduli space such that the periodic directions of moduli space are (sub-)Planckian in size. This means that, as in the 10d case of Type IIB considered above, any geodesic may be divided into a non-periodic variation of the dilaton/radion and a sub-Planckian periodic variation of the axion. The effect of the latter can be absorbed as an order-one shift of the coefficient $c_2$, whereas the former dominates for super-Planckian geodesic variation $|p-p_0| \gg \Mnine$, and our previous analysis ensures that conjecture 2 will once again be satisfied with $\lambda \geq 1/\sqrt{d-2}$.

This example also illustrates the problem with other possible refinements of the Distance Conjecture considered in Section \ref{sec:INTRO}. Let us once again ignore the axions and consider a geodesic in the $\phi$-$\rho$ plane of the form
\begin{equation}
(\phi(s), \rho(s)) = (-\sqrt{\frac{7}{32}} s, \frac{5}{\sqrt{32}} s - \rho_0)\,, ~~~~ s \in (-\infty, \infty)\,,
\label{geodmap}
\end{equation}
with $\rho_0 $ a constant. Note that this geodesic crosses multiple fundamental domains of the T/S-dualities of the theory, and the limit $s \rightarrow \infty$ is a strong coupling limit of Type IIB string theory. Nonetheless, BPS conditions ensure that the mass formulas in \eqref{scaling9d} apply even in the strong coupling regime, and the refined Distance Conjecture discussed in Section \ref{sec:INTRO} is generally held to apply to \emph{any} geodesic, irrespective of whether that geodesic lies in a single fundamental domain of the duality group.

Plugging \eqref{geodmap} into \eqref{scaling9d}, the three towers scale along the geodesic as
\begin{align}
\mKK = m_0 \exp(- \frac{s}{\sqrt{2}} +  \rho_0)\,,~~~\mwind = m_0 \exp( \frac{3s}{\sqrt{8}} - \rho_0) \,, ~~~
\mDwind &=  m_0  \exp(- \frac{\rho_0}{ 2} )\,,
\label{scalinggeo}
\end{align} 
 Thus, in the limit $s \rightarrow \infty$, the Kaluza-Klein modes become exponentially light, satisfying the Distance Conjecture with a coefficient of $\lambda = 1/\sqrt{2}$. In the limit $s \rightarrow - \infty$, the fundamental string winding modes become exponentially light, satisfying the Distance Conjecture with a coefficient of $\lambda = 3/\sqrt{8}$.

\begin{figure}
\begin{subfigure}{0.38\textwidth}
\includegraphics[width = 55mm]{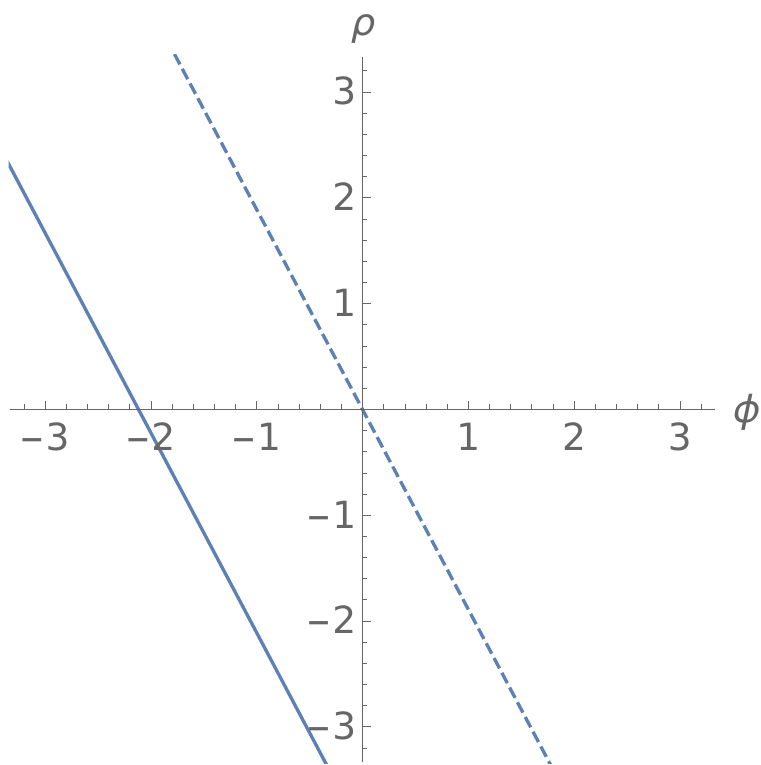} 
\end{subfigure}
\begin{subfigure}{0.3\textwidth}
\includegraphics[width = 53mm]{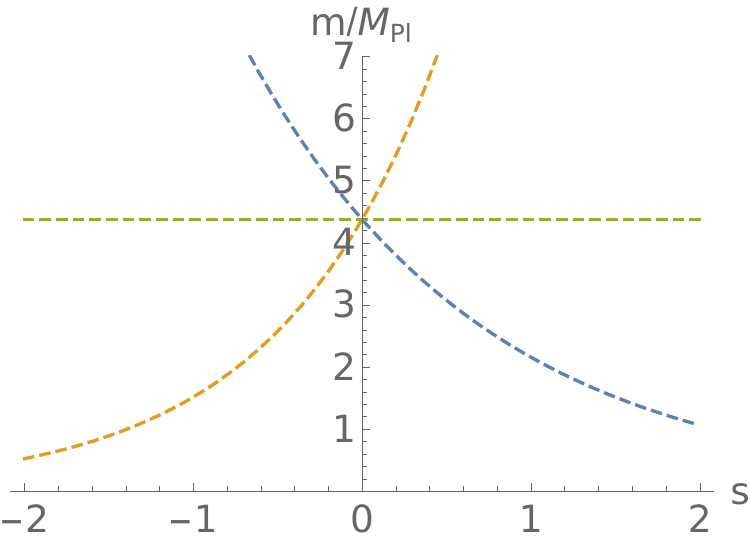} 
\end{subfigure}
\begin{subfigure}{0.3\textwidth}
\includegraphics[width = 53mm]{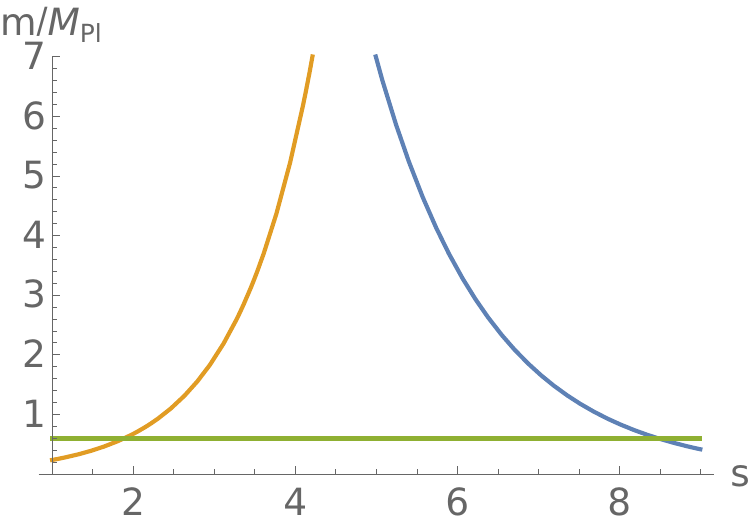} 
\end{subfigure}
\caption{Two geodesics in the dilaton-radion plane of Type IIB string theory on $S^1$ (left), and scaling of Kaluza-Klein modes (blue), winding modes (yellow), and D-string winding modes (green) along the dashed geodesic (middle) and the solid geodesic (right). By shifting the geodesic to the left in a parallel fashion, the point at which $\mKK = \mwind$ can be set arbitrarily larger than the Planck scale, and the range over which both of these scales are super-Planckian can be made arbitrarily large. Meanwhile, the D-string winding modes remain light everywhere along the geodesic.}
\label{geofig}
\end{figure}

In the intermediate regime, the relative behavior of the Kaluza-Klein modes and the fundamental string winding modes depends on the value of $\rho_0$. If $\rho_0 = 0$, then for $s >0$, the Kaluza-Klein modes are lighter, while for $s<0$ the winding modes are lighter. The crossover point occurs at $s=0$, where $\mKK = \mwind = m_0 \sim \Mnine$. 

For $\rho_0 \gg 1$, however, the crossover point between the Kaluza-Klein modes and the fundamental string winding modes occurs at an energy scale that is exponentially larger than the Planck scale, with $\mKK = \mwind \sim \exp(\rho_0/5) \Mnine$. Furthermore, both the Kaluza-Klein modes and the winding modes are heavier than the Planck scale in the range $3 \rho_0 /\sqrt{8} < s < \sqrt{2} \rho_0$. By taking $\rho_0$ large, therefore, we find a parametrically large range of length $\rho_0/\sqrt{8}$ over which both the Kaluza-Klein scale and the winding scale are parametrically heavier than the Planck scale. From this, we learn that the towers predicted from the Distance Conjecture in the asymptotic regimes $s \rightarrow \pm \infty$ do not necessarily play a role in the low-energy effective theory in the interior of moduli space, and indeed this interior region can be made parametrically large (see Figure \ref{geofig}).\footnote{Similar behavior was observed previously in certain AdS vacua in \cite{Conlon:2020wmc}.}

However, our analysis so far has not yet accounted for the D-string winding modes, whose masses are constant and sub-Planckian everywhere along the geodesic of interest for $\rho_0 > 0$. Once these are included, the minimum of $\mKK$, $\mwind$, and $\mDwind$ is bounded above by $m_0$ everywhere along the geodesic, and both of our proposed conjectures are satisfied.

\subsection{String theory on $T^{n}$, $n > 1$}

The analysis above can be generalized simply for toroidal compactification of multiple dimensions. Compactifying Type IIB string theory on $n$ circles of radius $R_i$, $i=1,...,n$, we find Kaluza-Klein modes, fundamental string winding modes, and D-string winding modes at the scales
\begin{equation}
\mKK^{(i)} = \frac{1}{R_i}\,,~~~\mwind^{(i)} = M_s^2 R_i \,,~~~~ \mDwind^{(i)} = \frac{M_s^2 R_i}{g_s}\,.
\end{equation} 
The Planck scale in $d = 10-n$ dimensions is given by 
\begin{equation}
M_{\textrm{Pl}; d} =  \left( \frac{4 \pi }{g_s^2}  \prod_{i=1}^n (2 \pi R_i) \right)^{1/(d-2)} \left( \frac{M_s}{2\pi} \right)^{8/(d-2)}
\end{equation}
Once again, the minimum of all of these scales is maximized when $R_i = 1/M_s$, $g_s = 1$. At this point in moduli space, which we denote $p_{\text{ctr}}$, we have for all $i$,
\begin{equation}
\mKK^{(i)} = \mwind^{(i)} = \mDwind^{(i)} = M_s = \frac{2 \pi \Mpld}{(4 \pi)^{1/(d-2)}}\,.
\end{equation}
There are also towers of particles associated with wrapped D$p$-branes for $p \leq n$ odd, and for $n \geq 5$ there is a tower associated with wrapped NS5-branes. At $p_{\text{ctr}}$, the mass scale associated with all of these towers is again $M_s$.

As in the previous subsection, we may define $\zeta$-vectors for the various towers of particles by \eqref{zeta}. By the analysis of \cite{Etheredge:2022opl}, the convex hull of these $\zeta$-vectors contains the ball of radius $1/\sqrt{d-2}$ centered at the origin. This means that as one moves in moduli space away from $p_{\text{ctr}}$, at least one of these towers becomes lighter in Planck units, with a mass that decays exponentially with $\lambda \geq 1/\sqrt{d-2}$. Any geodesic of length $L$ in radion-dilaton space will contain at least one point $p_1$ that lies a distance $L/2$ from $p_{\text{ctr}}$, and at this point at least one tower will have a mass that is bounded above by
\begin{equation}
m(p_1) \leq \frac{2 \pi \Mpld}{(4 \pi)^{1/(d-2)}} \exp(- L /(2\sqrt{d-2})  )\,.
\end{equation}
Thus, once again, our proposed conjectures are satisfied if the periodic directions of moduli space are ignored, with $c_1 = c_2 = 2 \pi /( 4 \pi)^{1/(d-2)}$, $\lambda = 1/\sqrt{d-2}$. By a similar argument to the 10d and 9d cases above, the inclusion of axions may alter $c_1$ and $c_2$ by an order-one amount, but it will not affect $\lambda$.

Our analysis here generalizes straightforwardly to Type IIA string theory on $T^n$, by T-duality. Furthermore, because our arguments relied on rather generic scaling behavior of Kaluza-Klein modes and wound strings, we expect that they will carry over to heterotic string heterotic string theories on $T^n$, and thus to M-theory/Type IIA string theory on K3, by duality. We will explore these scenarios further in a forthcoming work \cite{More}.

\section{M-theory on a Calabi-Yau Threefold}\label{5D}

In this section, we consider two simple yet illustrative examples of M-theory compactifications on Calabi-Yau manifolds. The geometries we consider are the symmetric flop geometry and the Greene-Morrison-Strominger-Vafa (GMSV) geometry \cite{Greene:1995hu, Greene:1996dh}, which were analyzed thoroughly in Section 7 of \cite{Alim:2021vhs}. Further details on these geometries (and 5d supergravity in general) can be found in that paper. 

We begin with the symmetric flop geometry. M-theory compactified on this geometry gives rise to a 5d supergravity theory with prepotential
\begin{equation}
\cF = \frac{1}{3} X^3 + 2 X^2 Y\,.
\end{equation}
Here, $X$ and $Y$ are real, and the vector multiplet moduli space is the one-dimensional slice given by $\cF = 1$.
The limit $X \rightarrow 0$ is an infinite-distance limit in moduli space, and it is accompanied by a tower of particles that satisfies the Distance Conjecture.

Meanwhile, a conifold singularity develops at the the point $Y=0$, and the geometry may be flopped to an isomorphic phase with 
\begin{equation}
\cF' = \frac{1}{3} (X')^3 + 2 (X')^2 Y'\,.
\end{equation}
Here, $X' = X + 4 Y$, and $Y' = -Y$. The physics is identical to that in the first phase, and once again there is a tower of particles that satisfies the Distance Conjecture in the infinite-distance limit $X' \rightarrow 0$, though it is not the same tower that satisfies the Distance Conjecture in the $X \rightarrow 0$ limit.

\begin{figure}
\begin{center}
\begin{subfigure}{0.45\textwidth}
\includegraphics[width = 70mm]{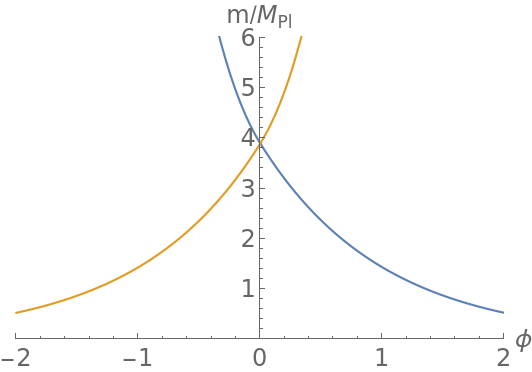} 
\end{subfigure}
\begin{subfigure}{0.45\textwidth}
\includegraphics[width = 65mm]{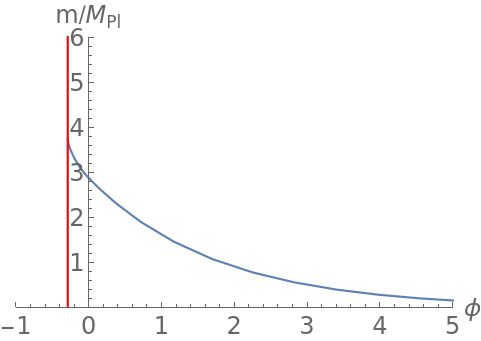} 
\end{subfigure}
\end{center}
\caption{Tower masses throughout the moduli space of the symmetric flop geometry (left) and the GMSV geometry (right). In the symmetric flop geometry, the limits $\phi \rightarrow \pm \infty$ each feature an exponentially light tower of particles, and the masses of these towers intersect at $\phi =0$ with $m =(6 \pi^2)^{1/3} \Mfive$. In the GMSV geometry, there is only one infinite-distance limit in vector multiplet moduli space, $\phi \rightarrow \infty$, but the mass of the tower which satisfies the Distance Conjecture in this limit is bounded above as $m \leq (16 \pi^2/3)^{1/3} \Mfive$.}
\label{5dfig}
\end{figure}

We may define a canonically normalized scalar field $\phi$, so that the $X \rightarrow 0$ limit corresponds to $\phi \rightarrow \infty$, the $X' \rightarrow 0$ limit corresponds to $\phi \rightarrow - \infty$, and the flop occurs at $\phi =0$. Then, the mass of the tower of particles that becomes light in the $X \rightarrow 0$ limit is given by \cite{Alim:2021vhs}
\begin{equation}
m(\phi) = \left\{
    \begin{array}{lr}
    (6 \pi^2)^{1/3} \Mfive  e^{- \phi /\sqrt{3} }  , & \text{if } \phi \geq 0\\
        (2 \pi^2 /9)^{1/3} \Mfive \left( 2 e^{ 2|\phi| /\sqrt{3}}+ e^{- |\phi| /\sqrt{3} }   \right)  , & \text{if } \phi < 0 
    \end{array}
\right\} \,.
\end{equation}
The mass of the tower of particles that becomes light in the $\phi \rightarrow - \infty$ limit is given by the same expression, but with $\phi \rightarrow - \phi$. The masses of these two towers are equal at the point $\phi = 0$, at which point we have $m(0) = (6 \pi^2)^{1/3} \Mfive$. For all other values of $\phi$, one of these towers is lighter than this value, as shown in Figure \ref{5dfig} (left). We see that conjecture 1 is satisfied everywhere in vector multiplet moduli space with a coefficient of $c_1 = (6 \pi^2)^{1/3}$.

Furthermore, any geodesic in vector multiplet moduli space of length $L$ will have at least point that lies a distance $L/2$ or greater from $\phi=0$, and correspondingly there will be a tower with mass $m(|\phi| = L/2) =   (6 \pi^2)^{1/3} \Mfive  \exp (- L /(2\sqrt{3}) )$. This ensures that conjecture 2 is satisfied with $c_2 = (6 \pi^2)^{1/3}$, $\lambda = 1/\sqrt{3} = 1/\sqrt{d-2}$.

The GMSV geometry similarly features one-dimensional vector multiplet moduli space, with prepotential given by
\begin{equation}
\cF = \frac{5}{6} X^3 + 2 X^2 Y\,.
\end{equation}
Once again, the limit $X \rightarrow 0$ is an infinite-distance limit in moduli space, and it is accompanied by a tower of particles that satisfies the Distance Conjecture. A conifold singularity develops at the the point $Y=0$, and the geometry may be flopped to another phase of the geometry. In this case, however, the second phase is not isomorphic to the first, but rather it features a prepotential of the form
\begin{equation}
\cF' = \frac{5}{6} (X')^3 + 8 (X')^2 Y'  + 24  X' (Y')^2 + 24 (Y')^3    \,,
\end{equation}
with $X' = X + 4 Y$, and $Y' = -Y$. In this case, the boundary $X' = 0$ lies at finite distance in moduli space.
This means that the vector multiplet moduli space is semi-infinite, and there is only one tower of BPS particles which satisfies the Distance Conjecture. In the limit $X \rightarrow 0$, this tower becomes exponentially light, whereas at the boundary $X' = 0$, this tower begins at the mass scale
\begin{equation}
m(X' =0) = (16 \pi^2/3)^{1/3} \Mfive\,.
\end{equation}
At all other points in moduli space, the mass of the tower is smaller than this value, as illustrated in Figure \ref{5dfig} (right).
Conjecture 1 is therefore satisfied with a coefficient of $c_1 = (16 \pi^2/3)^{1/3}$.

In the GMSV geometry, the mass of the lightest tower is maximized at a boundary of moduli space rather than in the interior of moduli space. From Figure \ref{5dfig} (right), one can see that the boundary $X' = 0$ (indicated by the vertical red line) is located a sub-Planckian distance $|\phi_{\text{min}}| \approx 0.295$ from the flop at $\phi = 0$. In the original, unflopped phase, the mass of the tower scales exponentially as 
\begin{equation}
m(\phi) = (12 \pi^2/5)^{1/3} \Mfive \exp(- \phi / \sqrt{3})\,,~~~~\text{if } \phi \geq 0\,,
\end{equation}
where $\phi$ is the geodesic distance from the flop at $\phi=0$. At a distance of $|\Delta \phi| = \phi + |\phi_{\text{min}}|$ from the boundary of moduli space, the mass of the tower is then bounded above as
\begin{equation}
m(|\Delta \phi|) \leq (16 \pi^2/3)^{1/3} \Mfive \exp( - |\Delta \phi | / \sqrt{3})\,, 
\end{equation}
so conjecture 2 is satisfied with $c_2 =  (16 \pi^2/3)^{1/3}$, $\lambda = 2 /\sqrt{3}$. The additional factor of two in $\lambda$ relative to the symmetric flop example is a consequence of the fact that the domain of $\phi$ is semi-infinite.

This example illustrates precisely the behavior required by the refined Distance Conjecture: the asymptotic exponential falloff of the tower mass kicks in at the flop, which lies a sub-Planckian distance from the boundary of moduli space. If (contrary to fact) the boundary of moduli space were to lie a super-Planckian distance from the flop, the refined Distance Conjecture could be violated, and furthermore the mass of the tower could rise high above the Planck scale. We see that, at least in this example, the refined Distance Conjecture and our proposed conjectures are closely related, as the boundary of moduli space is located an order-one distance from the point where tower mass is equal to the Planck mass and the exponential scaling of the tower mass kicks in.

M-theory compactifications on Calabi-Yau threefolds offer a vast playground for testing our proposed conjectures (and the refined Distance Conjecture) in greater detail. It would be worthwhile to explore this playground further.


\section{Einstein-Maxwell-Dilaton Theory in Four Dimensions}\label{EMD}

Consider an Einstein-Maxwell-dilaton theory in four dimensions with action
\begin{equation}
S = \frac{1}{2} \int d^4x \sqrt{-g} \left( \mathcal{R} - \frac{1}{2} (\nabla \phi)^2 - \frac{1}{g^2} e^{- \alpha \phi} F_2^2  \right) \,.
\end{equation} 
The tower Weak Gravity Conjecture \cite{Heidenreich:2015nta, Heidenreich:2016aqi, Andriolo:2018lvp} then predicts a tower of charged particles beginning at the mass scale\footnote{In this section, we define $\phi$, $g$ so that $\langle \phi \rangle =0$. A nonzero vev can be absorbed into the definition of $g$.}
\begin{equation}
m_{\text{el}} =g  \sqrt{\gamma}  \Mpl \,,
\end{equation}
where
\begin{equation}
\gamma^{-1} = \frac{1+ \alpha^2}{2}  \,.
\end{equation}
Far up in the tower, we expect that these states are extremal black holes \cite{Kats:2006xp}.

Meanwhile, the magnetic tower Weak Gravity Conjecture implies a tower of charged monopoles beginning at the scale 
\begin{equation}
m_{\text{mag}} = \frac{2 \pi  \sqrt{\gamma}}{g} \Mpl \,,
\end{equation}
where $\gamma$ is the same as the electric case. These two mass scales become equal to each other at the point $p_0$ in moduli space where $g = 2 \pi /g$, yielding 
\begin{equation}
m_{\text{el}}  = m_{\text{mon}} =   \sqrt{ 2 \pi \gamma}  \Mpl \leq 2 \sqrt{ \pi} \Mpl \,.
\end{equation}
We see, therefore, that for all values of $\phi$, either $m_{\text{el}}$ or $m_{\text{mon}}$ is parametrically at or below the Planck scale, and there exists a point at which each of these mass scales is equal to the Planck scale up to an order-one factor.
The coefficient $c_1$ is bounded above by $2 \sqrt{\pi}$, and it is smaller for larger values of $|\alpha|$.


\section{Discussion}\label{CONC}

In this note, we have sought to address one issue separating the Distance Conjecture from meaningful phenomenological constraints by exploring the extent to which the Distance Conjecture constrains effective field theories in the interior of moduli space. We saw that in some cases, the refined Distance Conjecture is satisfied only by super-Planckian towers of particles, limiting its ability to constrain low-energy physics. However, in all cases we considered, at least one tower of particles begins at or below the Planck scale, and for a long geodesic path it is exponentially lighter than the Planck scale, a phenomenon which we codified in a pair of closely related conjectures.

We showed how these conjectures are satisfied in number of examples in string/M-theory. In theories with sufficient supersymmetry, the conjectures are satisfied because multiple towers coalesce near the Planck scale. The prototypical example of this is a circle compactification, in which a tower of Kaluza-Klein modes becomes light in the large-radius limit, a tower of winding modes becomes light in the small-radius limit, and at the self-dual radius these towers coalesce. T-duality exchanges the Kaluza-Klein and winding modes, ensuring that the Emergent String Conjecture is satisfied in the small-radius limit.

In other theories, however, there is an alternative possibility: the minimum tower mass might obtain its maximum at a finite-distance boundary of moduli space, as in the 5d GMSV example considered in Section \ref{5D} above. In this case, our conjectures suggest that the finite-distance boundary should be close (in Planck units) to the locus in moduli space where the lightest tower has Planck-scale mass. In the GMSV example, this locus was also close to the locus where the exponential scaling behavior of the tower mass set in, resulting in a close connection between our proposed conjectures and the original refined Distance Conjecture.

Further exploration of string compactifications could lead to further evidence for or against our conjectures. As noted previously, one promising arena for future study is the vector multiplet moduli spaces of 5d theories arising from M-theory compactified on Calabi-Yau threefolds. Another intriguing class of examples is the
small-volume limits of Type II string theory on Calabi-Yau threefolds. Such limits may be classically at infinite distance, but quantum corrections to the metric cut off the moduli space at finite distance \cite{Marchesano:2019ifh,Baume:2019sry, Klaewer:2020lfg}. In this case, conjecture 1 would seem to require that the quantum corrections must obstruct the portion of moduli space in which the Kaluza-Klein scale is heavier than the Planck scale, and it would be interesting to see how this plays out in practice.

Even if our conjectures are true, there are still several steps separating the Distance Conjecture in its current form from meaningful constraints on inflation. For one thing, the strongest evidence in favor of the Distance Conjecture comes from the supersymmetric context, where the scalar fields in question are massless moduli. The idea that the Distance Conjecture should apply more generally to non-supersymmetric theories and scalar fields with potentials goes all the way back to the original paper on the Distance Conjecture by Ooguri and Vafa \cite{Ooguri:2006in}, and while the evidence so far seems consistent with this idea, further analysis is needed.
%

It is also not entirely clear that a tower of particles at a mass scale $m < H$ implies a problem for inflation at the Hubble scale, $H$. Here, the Emergent String Conjecture \cite{Lee:2019wij} becomes relevant, as this conjecture implies that any infinite-distance limit in moduli space is either a decompactification limit or an emergent string limit. Combined with the analysis of \cite{Etheredge:2022opl}, this suggests that the lightest tower in any infinite-distance limit should be either a Kaluza-Klein tower or an emergent string tower in some duality frame. For such a tower, there \emph{is} a problem with $m < H$: if the Kaluza-Klein scale drops below the Hubble scale, the theory experiences decompactification and cannot be considered a $4$-dimensional FRW cosmology. If the string scale drops below the Hubble scale, the universe enters a stringy phase, and low-energy effective field theory no longer applies.

Like the Distance Conjecture, it is not clear that the Emergent String Conjecture applies in the non-supersymmetric context to scalar fields with potentials. Preliminary evidence suggests that this may be true \cite{Marchesano:2019ifh, Baume:2019sry, Klaewer:2020lfg, Basile:2022zee}, but more research is needed.

Finally, our analysis has led us to a bound on tower masses as a function of geodesic distance between two points, but the path traversed by the inflaton is not necessarily a geodesic. One prominent example of this is axion monodromy inflation \cite{Silverstein:2008sg,McAllister:2008hb}, in which an axion winds around its fundamental domain multiple times. Further work is needed to understand the quantum gravity constraints on these models of inflation.

In summary, the gap between the Distance Conjecture and robust phenomenological constraints is still formidable, though it has narrowed in recent years. Further research into the validity and scope of the conjectures proposed here could play an important role in narrowing this gap further.



\section*{Acknowledgements}

It is a pleasure to thank Ben Heidenreich for useful discussions. This work was supported in part by the Berkeley Center for Theoretical Physics; by the Department of Energy, Office of Science, Office of High Energy Physics under QuantISED Award DE-SC0019380 and under contract DE-AC02-05CH11231; and by the National Science Foundation under Award Number 2112880.

\bibliographystyle{utphys}
\bibliography{ref}

\end{document}